\documentclass[twocolumn]{aastex631}

\usepackage{CJK}

\usepackage{amsmath}
\usepackage{booktabs}
\usepackage{multirow}

\newcommand{\blue}{\color{blue}}

\newcommand{\beq}	{\begin{equation}}
\newcommand{\eeq}	{\end{equation}}

\def\f0{\ifmmode {f_{\rm V,0}} \else $f_{\rm V,0}$\fi}
\def\fv{\ifmmode {f_{\rm V}} \else $f_{\rm V}$\fi}
\def\nh{\ifmmode {n_{\rm H}} \else $n_{\rm H}$\fi}
\def\n0{\ifmmode {n_{\rm H,0}} \else $n_{\rm H,0}$\fi}
\def\fhi{\ifmmode {f_{\rm HI}} \else $f_{\rm HI}$\fi}
\def\Teq{\ifmmode {T_{\rm eq}} \else $T_{\rm eq}$\fi}
\def\nhi{\ifmmode {N_{\rm HI}} \else $N_{\rm HI}$\fi}
\def\ni0{\ifmmode {N_{\rm HI,0}} \else $N_{\rm HI,0}$\fi}

\def\rvir{\ifmmode {R_{\rm vir}} \else $R_{\rm vir}$\fi}
\def\mvir{\ifmmode {M_{\rm vir}} \else $M_{\rm vir}$\fi}
\def\r200{\ifmmode {R_{\rm 200m}} \else $R_{\rm 200m}$\fi}
\def\m200{\ifmmode {M_{\rm 200m}} \else $M_{\rm 200m}$\fi}
\def\rcgm{\ifmmode {R_{\rm CGM}} \else $R_{\rm CGM}$\fi}
\def\mcgm{\ifmmode {M_{\rm CGM}} \else $M_{\rm CGM}$\fi}

\def\mcgmcool{\ifmmode {M_{\rm cCGM}} \else $M_{\rm cCGM}$\fi}
\def\fcgmcool{\ifmmode {f_{\rm cCGM}} \else $f_{\rm cCGM}$\fi}
\def\mbcosm{\ifmmode {M_{\rm b,cosm}} \else $M_{\rm b,cosm}$\fi}

\def\mmet{\ifmmode {M_{\rm metals}} \else $M_{\rm metals}$\fi}
\def\mstar{\ifmmode {M_{*}} \else $M_{*}$\fi}
\def\tdyn{\ifmmode {t_{\rm dyn}} \else $t_{\rm dyn}$\fi}

\def\mp{\ifmmode {m_{\rm p}} \else $m_{\rm p}$\fi}
\def\cmv{\ifmmode {\rm cm^{-3}} \else ${\rm cm^{-3}}$\fi}
\def\cmc{\ifmmode {\rm cm^{-2}} \else ${\rm cm^{-2}}$\fi}
\def\msun{\ifmmode {M_{\rm \odot}} \else $M_{\rm \odot}$\fi}

\def\dmcool{\ifmmode {\dot{M}_{\rm cool}} \else $\dot{M}_{\rm cool}$\fi}
\def\dmacc{\ifmmode {\dot{M}_{\rm acc}} \else $\dot{M}_{\rm acc}$\fi}
\def\msuny{\ifmmode {\rm  M_{\odot}~{\rm yr^{-1}}} \else ${\rm M_{\odot}~{\rm yr^{-1}}}$\fi}

\def\imppar{\ifmmode {r_{\perp}} \else $r_{\perp}$\fi}

\newcommand{\HI}{H{\sc\ i}}

\newcommand{\SiII}{Si{\sc\ ii}}

\newcommand{\SiIV}{Si{\sc\ iv}}

\newcommand{\CIV}{C{\sc\ iv}}
\newcommand{\OVI}{O{\sc\ vi}}

\defcitealias{Prochaska17}{P17}
\defcitealias{FW23}{FW23}
\defcitealias{KS19}{KS19}
\defcitealias{Faerman24}{F24}
\defcitealias{Zheng24}{Z24}
\defcitealias{Mishra24}{M24}

\shorttitle{Limits on Cool CGM in Dwarfs}
\shortauthors{Faerman, Zheng, \& Oppenheimer}

\begin{document}

\begin{CJK*}{UTF8}{gbsn} 

\title{Upper Limits on the Mass of Cool Gas in the Circumgalactic Medium of Dwarf Galaxies}

\author[0000-0003-3520-6503]{Yakov Faerman}
\affiliation{Department of Astronomy, University of Washington, Seattle, WA 98195, USA, {\blue yakov.faerman@gmail.com}} 

\author[0000-0003-4158-5116]{Yong Zheng (郑永)}
\affiliation{Department of Physics, Applied Physics and Astronomy, Rensselaer Polytechnic Institute, Troy, NY 12180, USA}

\author[0000-0002-3391-2116]{Benjamin D. Oppenheimer}
\affiliation{CASA, Department of Astrophysical and Planetary Sciences, University of Colorado, Boulder, CO 80309, USA}

\begin{abstract}

We use HI absorption measurements to constrain the amount of cool ($\approx 10^4$~K), photoionized gas in the circumgalactic medium (CGM) of dwarf galaxies with $M_* = 10^{6.5-9.5}~\msun$ in the nearby Universe ($z<0.3$). We show analytically that volume-filling gas gives an upper limit on the gas mass needed to reproduce a given HI column density profile. We introduce a power-law density profile for the gas distribution and fit our model to archival HI observations to infer the cool CGM gas mass, $\mcgmcool$, as a function of halo mass. For volume-filling ($\fv=1$) models, we find $\mcgmcool = 5 \times 10^8-2 \times 10^9$~\msun, constituting $\lesssim 10\%$ of the halo baryon budget. For clumpy gas, with $\fv=0.01$, the masses are a factor of $\approx 11$ lower, in agreement with our analytic approximation. Our assumption that the measured HI forms entirely in the cool CGM provides a conservative upper limit on $\mcgmcool$, and possible contributions from the intergalactic medium or warm/hot CGM will further strengthen our result. We estimate the mass uncertainties due to the range of redshifts in our sample and the unknown gas metallicity to be $\approx 15\%$ and $\approx 10\%$, respectively. Our results show that dwarf galaxies have only $\lesssim 15\%$ of their baryon budget in stars and the cool CGM, with the rest residing in the warm/hot CGM or ejected from the dark matter halos.
\end{abstract}



\section{Introduction} \label{sec:intro}

The circumgalactic medium (CGM) connects the large-scale structure and the galaxies that reside at the nodes of the cosmic web. Gas accreted from the intergalactic medium (IGM) and stripped from satellite galaxies is deposited into the CGM, which can then fuel star formation and the growth of supermassive black holes \citep[e.g.,][]{faucher-giguere23}. Galactic processes, such as stellar and active galactic nuclei (AGN) feedback, in turn shape the CGM by heating it up and enriching it with metals \citep[e.g.,][]{crain23}. The last decades saw significant progress in multiwavelength observations, revealing that the CGM is a massive, complex, and multiphase structure \citep{tumlinson17}.

Empirical theoretical studies of the physical state of the CGM have largely focused on Milky Way-mass and more massive galaxy halos \citep[e.g.,][]{MB04, Stern16, Stern18, FSM17, mcquinn18, Qu18a, Qu18b, Afruni19, Afruni21, LanMo19, FW23}. Such models allow relating the gas physical distributions to the various CGM observables, which in turn helps constrain the properties of the CGM, such as the gas mass and metallicity, gas and energy flows, and small-scale physics (see also recent comparison by \citealp{Singh24}). Meanwhile, the CGM of dwarf galaxies remains largely uncharacterized, and it is unclear whether the baryon cycles in dwarf galaxies are similar to or different from their more massive counterparts.

Dwarf galaxies are great laboratories for understanding star formation and feedback processes \citep{sales22}. With their shallower gravitational potentials, feedback in dwarf galaxies can be more efficient in transporting a majority of the produced metals into the CGM and beyond \citep{maclow99, tremonti04, kirby13, collins22}. Indeed, existing observations have shown that the CGM of nearby and low-$z$ dwarf galaxies can be detected in various ions in the ultraviolet (UV) such as \HI, \SiII, \SiIV, \CIV, and \OVI, although the detection rates of metal ions drop significantly at large impact parameters (\citealt{Bordoloi14, liang14, Johnson17, KT22, Zheng24}).

Recent work by \citet[][hereafter \citetalias{Zheng24}]{Zheng24} provided a compilation and comprehensive analysis of archival  (\citealp{Bordoloi14, Bordoloi18, liang14, Johnson17, Zheng20, QuBregman22}) and new absorption measurements of hydrogen and metals detected in the CGM of dwarf galaxies with stellar masses in the range $10^{6.5}-10^{9.5}~\msun$, corresponding to a halo mass range of $\m200 \approx 10^{10-11.5}~\msun$\footnote{$\m200$ is the mass enclosed by a radius inside which the mean matter density is a factor of 200 higher than that of the Universe.}. More recently, \citet[][hereafter \citetalias{Mishra24}]{Mishra24} reported a new set of absorption HI and metal absorption measurements of low-mass galaxies with a similar range of stellar masses as part of the CUBS survey (see also \citealp{Chen20}). \citetalias{Zheng24} estimated for their sample that the total mass of cool CGM gas, $\mcgmcool$, typically traced by UV ions only constitutes $\sim 2\%$ of the total baryon budget with $\mcgmcool=10^{8.4}~\msun$. However, it is unclear whether the cool-gas mass is constant across the halo mass range probed by \citetalias{Zheng24}, or there is a significant difference between low- and high-mass dwarf galaxy halos.

In this work, we construct a phenomenological CGM model, apply it to a sample combining data from \citetalias{Zheng24} and \citetalias{Mishra24}, and investigate how the cool CGM mass varies as a function of dwarf galaxy halo mass. This manuscript is organized as follows. In Section~\ref{sec:model} we review our data set, present our model for the gas distribution, and infer upper limits on $\mcgmcool$. We then discuss the implications of our results for dwarf galaxies' CGM gas properties and galaxy structure and evolution in Section~\ref{sec:disc} and summarize in Section~\ref{sec:summary}.

\section{Models and Results} \label{sec:model}

We now construct a model for the distribution of cool gas ($T\approx 10^4$~K) in the CGM and apply it to observations to infer an upper limit on the cool CGM gas mass, $\mcgmcool$. 
We first show an idealized case to demonstrate that volume-filling gas requires the largest mass needed to reproduce a given HI column density, assuming photoionization equilibrium (PIE). We then present our detailed model for the gas density distribution, temperature, and ionization. We apply the model to the data to derive an upper limit for $\mcgmcool$ as a function of halo mass and discuss its uncertainties.

\subsection{Data set}\label{subsec:model_data}

We use the dwarf galaxy sample presented in \citetalias{Zheng24}, residing at redshifts between $z=0$ and $0.3$. We add to this sample objects from the recent study by \citetalias{Mishra24}, which extends to higher redshifts, and we select galaxies at $z<0.3$ for modeling consistency (see below). Given the reported galaxy stellar masses, in the range $10^{6.5-9.5}$~\msun~(see Figure~1 in \citetalias{Zheng24}), we use the UNIVERSE MACHINE tool kit \citep{Behroozi19} to convert stellar mass to halo mass, and the COLOSSUS toolkit \citep{Diemer18} to convert to $M_{200m}$ and $R_{200m}$ for consistency with \citetalias{Zheng24}. The range of stellar masses in the sample corresponds to halo masses of $\m200 \approx 10^{10.0-11.5}~\msun$, with $\r200 \approx 70-210$~kpc. For the reported stellar mass uncertainties, $0.1-0.2$ dex, the typical uncertainties for $\m200$ and $\r200$ are $0.1$ and $0.03$ dex, respectively.

We focus on galaxy-quasar pairs with HI observations and normalized impact parameters, $\imppar/\r200$, below unity. This results in $25$ objects from the \citetalias{Zheng24} sample, and $15$ objects from \citetalias{Mishra24}. We bin our sample by halo mass into three bins with similar number of objects. This results in $\log(\m200/\msun)$ bins of $10.20-10.80$, $10.80-11.15$, and $11.15-11.50$, containing $12$, $14$, and $14$ dwarf galaxies, respectively. From now on we refer to these as the low-, mid- and high-mass bins, and the median halo mass in each bin is given in Table~\ref{tab:params}.

While the individual galaxy redshifts are known, binning them by halo mass requires to choose a single redshift for each bin. We perform our modeling assuming $z=0$ for easier comparison between halo masses, interpretation of our results, and for consistency with \citetalias{Zheng24}, and we discuss the effect of redshift on our results in \S\ref{subsec:model_uncert}. The median galaxy redshift in our sample is $z=0.1$, and $90\%$ of the galaxies are at $z<0.2$.

We list the measured galaxy properties and the HI columns in Table~\ref{tab:data}. We plot the HI columns, $\nhi$, vs. normalized impact parameters, $\imppar/\r200$, in the right panel of Figure~\ref{fig:thermal_gen} (all data) and in the top panels of Figure~\ref{fig:col_data} (binned by halo mass, including the uncertainties in columns and impact parameters).

\subsection{The mass of cool, HI-traced gas is bound}\label{subsec:model_analytic}

We now show that for a measured HI column density, the (photoionized) gas mass needed to reproduce it is maximized for volume-filling gas. We describe an idealized case and show the main results here, and the detailed derivation is given in Appendix~\ref{app_analytic}.

For constant-density gas, the neutral hydrogen column density can be written as $\nhi = \nh \fhi L \fv$ where $\nh$ is the hydrogen volume density, $\fhi$ is the neutral hydrogen fraction, $L$ is the path length along the line of sight, and $\fv$ is the cool-gas volume filling fraction. For a given line of sight, $L$ is a constant, depending of the extent of the CGM and the projected distance from the galaxy. At low densities the neutral fraction can be approximated as a power-law function of the density, $\fhi \propto n_{\rm H}^\beta$. For example, for gas at $T=10^4$~K and densities in the range $10^{-6} < \nh/\cmv < 10^{-2}$, $\beta \approx 1$, and we can write
\beq\label{eq:nhi_prop}
\nhi \propto n_{\rm H}^2 \fv ~~~.
\eeq

The total cool-gas mass is given by $\mcgmcool = \frac{4\pi}{3} \bar{m} r_{\rm CGM}^3 \nh \fv$. Solving Equation~\eqref{eq:nhi_prop} for the gas density or the volume filling fraction and inserting into the expression for the gas mass gives
\beq\label{eq:mass_prop}
\mcgmcool \propto \nhi n_{\rm H}^{-1} \propto N_{\rm HI}^{1/2} f_{\rm V}^{1/2} ~~~,
\eeq
and since $\fv \leq 1$, volume-filling gas leads to an upper limit for $\mcgmcool$.

In reality, the gas density and volume filling fraction may vary as a function of radius, and the cool gas temperature is a function of the gas density, affecting the value of $\beta$ (see \S\ref{subsec:model_full} and Appendix~\ref{app_analytic}). These details do not change the main result of Equation~\eqref{eq:mass_prop}: For a given HI column density or column density profile, photoionized gas at lower densities is more ionized, hence the gas mass increases with the volume filling fraction, and is maximized for $\fv=1$. In the next section we present our detailed model and show that Eq.~\eqref{eq:mass_prop} provides a good approximation for the full numerical calculation.

\subsection{Detailed cool gas model}\label{subsec:model_full}

\begin{figure*}
\centering
\includegraphics[width=0.99\textwidth]{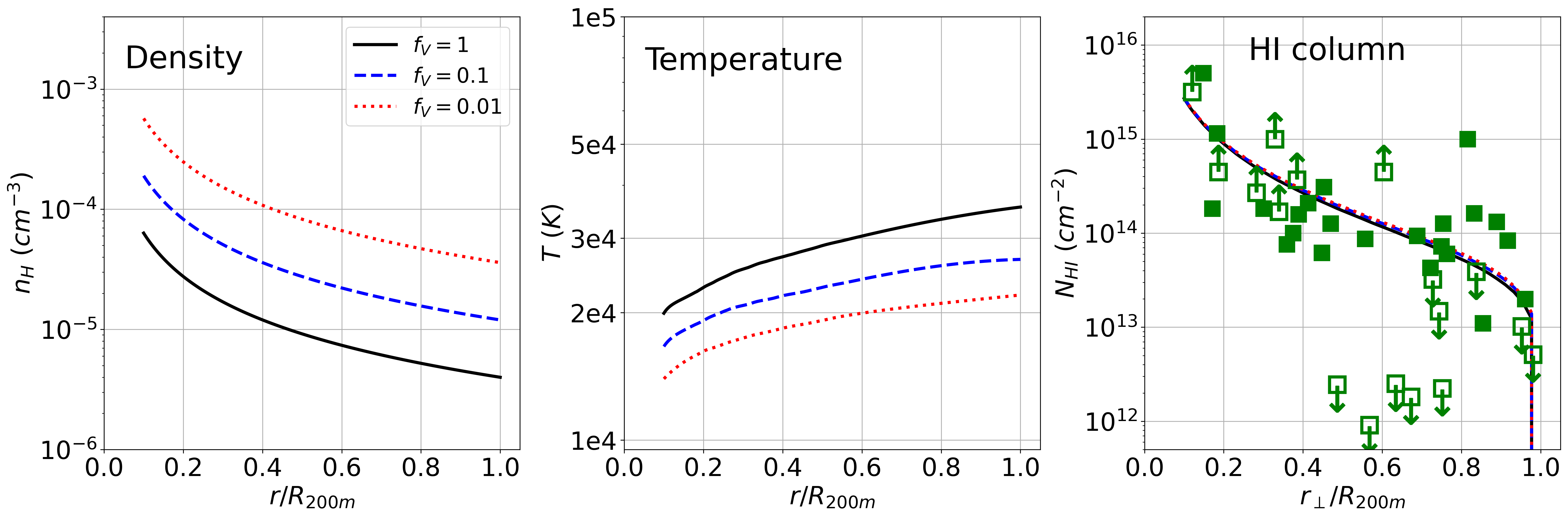}
\caption{The effect of the volume filling fraction on the gas properties for a given HI column density profile (and fixed density slope, $a_n$). The three models have $\fv$ between unity (solid black) and $0.01$ (dotted red) and the gas total mass varies to produce the same HI column density profile (right panel), roughly following the measured column densities (markers). Volume-filling gas (solid black curve) has lower volume densities (left panel) and higher temperatures (middle) (see text in \S\ref{subsec:model_full}).}\label{fig:thermal_gen}
\end{figure*}

Our goal in this work is to construct a minimal model needed to reproduce the observed HI column density profile. The main feature of this model is a power-law distribution of the gas density as a function of (radial) distance from the central galaxy,
\beq\label{eq:nhr}
\nh(r) = \n0\left(\frac{r}{\rcgm}\right)^{-a_n} ~~~,
\eeq
where we take $\rcgm$ to be $\r200$. The same density distribution is assumed in \citetalias{Zheng24} (see their Section~4), and other works have either assumed similar profiles \citep[e.g.,][]{Anderson13}, or shown that it is a good approximation to more detailed physical models or simulation results (see \citealp{F22,Oren24} for the warm/hot CGM, for example, or \citealp{FW23} for the cool CGM). In this work we also follow the choice made by \citetalias{Zheng24} for the extent of the CGM, and assume it resides between $0.1$ and $1$ $\r200$.

We make two changes from the model presented in \citetalias{Zheng24}. First, we assume that the gas is at a heating/cooling equilibrium with the metagalactic (UV) background radiation field (MGRF), and its temperature is~\Teq, set by the gas density and metallicity. This is different from the constant value $T=10^4$~K used in \citetalias{Zheng24} and often adopted in ionization modeling, and accounts for the effect of a self-consistent temperature on the gas ionization state. We use Cloudy~C22 \citep{Ferland17,Chatzikos23} to calculate \Teq~for a range of densities and metallicities, adopting the \cite{KS19} $z=0$ standard MGRF. As shown in \citet[][hereafter \citetalias{Faerman24}]{Faerman24}, for metallicities in the range $0.1-1.0$~solar, the gas temperature is a weak function of the metallicity (see their Figure~1 and Equation~12). In this work we adopt a nominal value of $Z'=0.3$, as done by \citetalias{Zheng24} (see also \citealp{Prochaska17}), and calculate the uncertainty this assumption introduces on our results\footnote{Since we only examine HI absorption, the metallicity does not have a direct impact on our models other than through this weak effect on the temperature.}. Finally, while the temperature is also only a weak function of the gas density (see \S\ref{subsec:disc_pressure}), it can vary by a factor of a few over the range of densities in our models (see middle panel of Figure~\ref{fig:thermal_gen}).

Second, we assume that the cool-gas volume filling fraction is constant with radius. This is motivated by the fact that variation in the density profile slope is enough to reproduce the observed HI column densities, and we have no empirical information about the behavior of $\fv$ as a function of distance from the galaxy. This simplification makes our model more robust and, as we discuss in \S\ref{subsec:model_results} and in Appendix~\ref{app_analytic}, our main results are not affected by this assumption.

To calculate the HI column densities, we use Cloudy to compute the neutral hydrogen fraction as a function of gas density and temperature, covering a wide range of possible conditions. In our Cloudy models we assume optically thin gas, a valid assumption given the low HI column densities measured around dwarf galaxies, with $\nhi \lesssim 10^{16}$~\cmc~(see discussion in Section~4.3 of \citetalias{Faerman24}). Given the gas radial distributions, we construct the HI fraction versus radius and calculate the HI column density profile for an external observer as a function of the impact parameter, $\imppar$.

Before we apply our model to the observations, we briefly demonstrate how the gas properties vary for different values of $\fv$ and $\nh$ for a fixed value of the density slope, $a_n$, producing similar HI column density profiles (see Section~4.2 in \citetalias{Zheng24} for a similar exercise). Figure~\ref{fig:thermal_gen} shows the gas densities (left panel), temperatures (middle), and the resulting HI column densities (right) for three models, with $\fv=1$ (solid black), $0.1$ (dashed blue), and $0.01$ (dotted red). For these models, we use a halo with $\r200 = 136$~kpc (mid-mass bin). All three models are constructed with $a_n=1.2$, and since they are tuned to produce the same HI column density, the density normalization increases for lower $\fv$, spanning a factor of $\approx 11$ and scaling approximately as $\n0 \propto f_{\rm V}^{-1/2}$ (see Eq.~\ref{eq:nhi_prop}). The gas temperature is set by the MGRF and is higher for lower gas densities, leading to a temperature increase with radius. The overall variation in temperature for a given model is a factor of $\approx 2$, with  a total range between $10^4$ and $4 \times 10^4$~K for all three models. The resulting gas masses are $1.0 \times 10^{9}$ (for $\fv=1)$, $3.0 \times 10^{8}$, and $9.0 \times 10^{7}$~\msun~($\fv=0.01$), similar to the approximation given in Eq.~\eqref{eq:mass_prop}.

\subsection{Applying the model to the data} \label{subsec:model_results}

\begin{figure*}
\centering{
\includegraphics[width=0.99\textwidth]{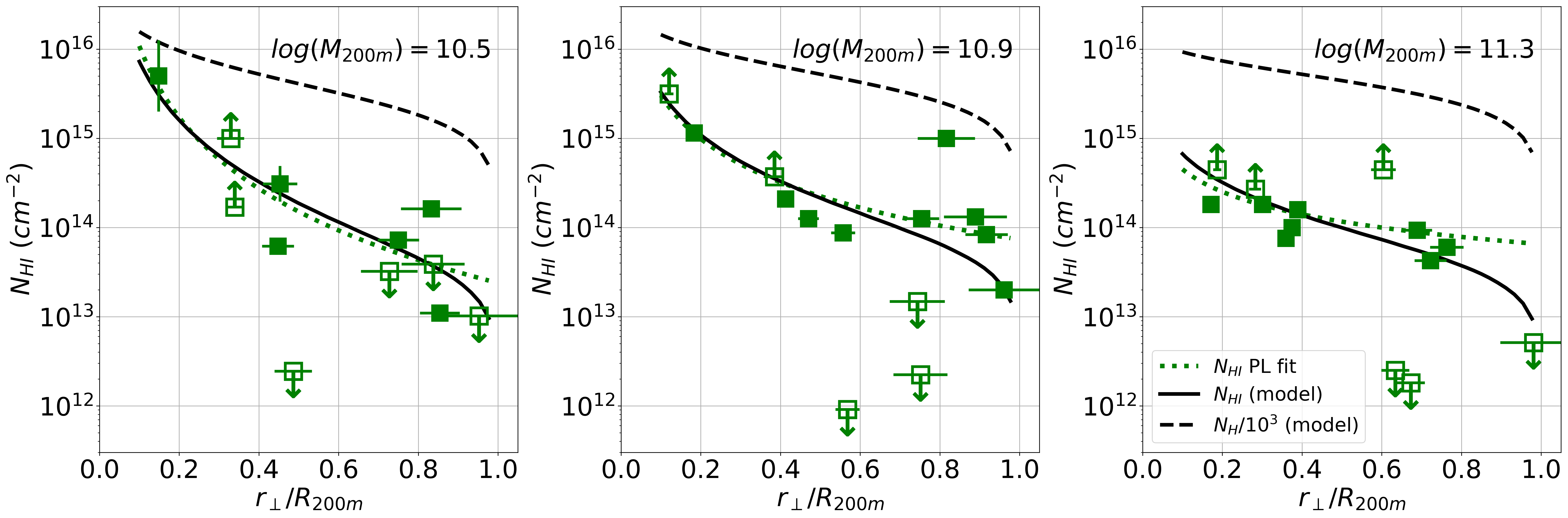}
\includegraphics[width=0.99\textwidth]{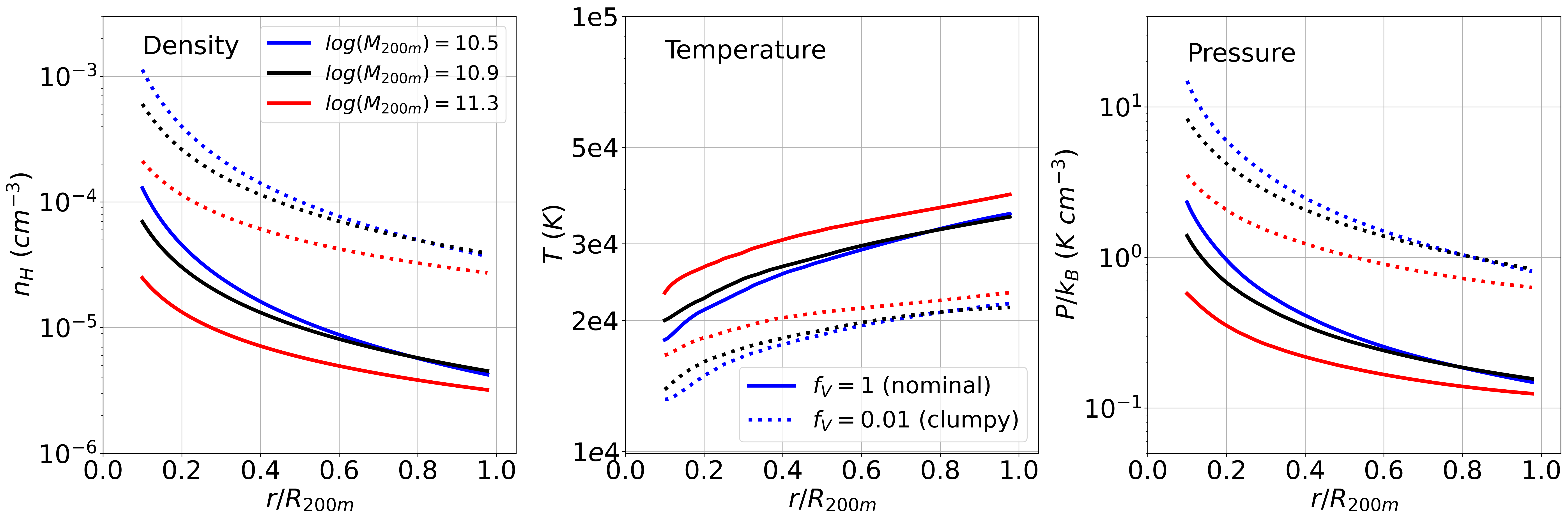}}
\caption{Our models fitted to the HI column densities measured in absorption, after binning the data in halo mass. 
{\bf Top: comparison to observations.} Markers: observed HI column densities (full and empty markers show measurements and limits, respectively), dotted curves are power-law fits to the column density profile to guide the eye (upper limits are ignored for this fitting), and solid black curves are our models. Dashed curves show the model total hydrogen columns (scaled down by a factor of $10^3$) for the volume-filling ($\fv=1$) scenario.
{\bf Bottom: gas thermal properties.} Density (left), temperature (middle), and pressure (right). The blue, black, and red curves correspond to the halo mass bins, from low to high, respectively. Solid and dotted curves are for the nominal and clumpy models, respectively (see \S\ref{subsec:model_results} and Table \ref{tab:params} for model description and parameters).}\label{fig:col_data}
\end{figure*}

We now fit the model presented in \S\ref{subsec:model_full} to the HI column density observations described in \S\ref{subsec:model_data} to estimate the cool CGM mass in the galaxies in our sample. The data binned by halo mass are shown in the top panels of Figure~\ref{fig:col_data}, with the measurements shown by the green squares, and the limits - by the empty squares with upward- and downward-pointing arrows. The green dotted curves show power-law fits to the data, to guide the eye. We fit our model by fixing $\fv$ and tuning the two density distribution parameters - the slope, $a_n$, and the normalization, $\n0$ - to fit the observed HI column densities in each bin. To do this, we construct models with different slopes and normalizations, calculate $\nhi(\imppar)$, and compare these to the measurements\footnote{For this calculation, we assume that the observed $\nhi$ originate entirely in the cool, photoionized CGM, providing a conservative limit for the gas mass in this phase. We examine possible contributions from the IGM or the warm/hot CGM in \S\ref{subsec:disc_igm_cont}.}. We then integrate over the density profiles to obtain \mcgmcool, the total cool gas mass between $0.1 \leq r/\r200 \leq 1$. We now discuss our fits to the data, and the parameters for the models described in this section are given in Table~\ref{tab:params}.

\begin{table*}
\centering
 	\caption{Model parameters and outputs}
 	\label{tab:params}
\begin{tabular}{| l || c | c | c | c | c | c | c | c | c | c |}
\toprule
Model set                           & $\r200$~(kpc) & $\m200$~(\msun) & $\fv$  & $a_n$ & $\n0$~(\cmv) & $\mcgmcool~(\msun)$ & $\frac{f_{\rm b,cool}}{f_{\rm b,cosm}} ($\%$)$ & $M_{\rm HI}~(\msun)$ & $Z'$ & $\mmet~(\msun)$ \\
\midrule
\multirow{3}{*}{\bf Nominal}

& $100$ & $3.2 \times 10^{10}$ & \multirow{3}{*}{$1$} & $1.5$ & $4.1 \times 10^{-6}$ & $4.8 \times 10^8$& $9.6$  & $7.3 \times 10^4$  & \multirow{3}{*}{$0.3$} & $2.1 \times 10^6$  \tabularnewline
& $136$ & $7.9 \times 10^{10}$ &                      & $1.2$ & $4.4 \times 10^{-6}$ & $1.1 \times 10^9$& $8.7$  & $1.1 \times 10^5$  & & $4.7 \times 10^6$  \tabularnewline
& $185$ & $2.0 \times 10^{11}$ &                      & $0.9$ & $3.1 \times 10^{-6}$ & $1.7 \times 10^9$& $5.4$  & $7.3 \times 10^4$  & & $7.3 \times 10^6$  \tabularnewline
\midrule
\multirow{3}{*}{\bf Clumpy}
& $100$ & $3.2 \times 10^{10}$ & \multirow{3}{*}{$0.01$} & $1.5$ & $3.6 \times 10^{-5}$ & $4.2 \times 10^7$ & $0.84$ & $7.2 \times 10^4$ & \multirow{3}{*}{$0.3$} & $1.8 \times 10^5$ \tabularnewline 
& $136$ & $7.9 \times 10^{10}$ &                        & $1.2$ & $3.8 \times 10^{-5}$ & $9.5 \times 10^7$ & $0.75$ & $1.0 \times 10^5$ &  & $4.1 \times 10^5$ \tabularnewline
& $185$ & $2.0 \times 10^{11}$ &                        & $0.9$ & $2.7 \times 10^{-5}$ & $1.4 \times 10^8$ & $0.46$ & $7.4 \times 10^4$ &  & $6.2 \times 10^5$ \tabularnewline
\bottomrule
\end{tabular}
\end{table*}

The HI column density profiles from our models are shown by the solid black curves in the top panels of Figure~\ref{fig:col_data}. The bottom panels show the density, temperature, and pressure profiles as functions of radius. For our nominal models we assume $\fv=1$, and these are shown by the solid curves (dotted curves are for models with $\fv=0.01$, which we discuss below). At densities below $10^{-2}$~\cmv, the neutral hydrogen fraction is low (see Appendix~\ref{app_analytic}), and the gas is mostly ionized. The dashed black curves in the top panels show the total hydrogen columns, scaled by a factor of $10^3$.

The observed HI columns in a given mass bin and impact parameter have some scatter, and may include lower and upper limits. These can come from variation in the cool gas properties between different galaxies or patchiness of the CGM in a single galaxy (see discussion in \citetalias{FW23} and in \citealp{SB24-mcc}). In this work we implicitly assume the former scenario and a sky-covering fraction of unity for the cool CGM gas in a given object, consistent with the nondetections constituting a small fraction of the observations. The fits we present are consistent with most of the measurements in each bin, and are close to many of the lower limits. Significantly higher lower limits are few and may not be representative of the average galaxy we aim to model (see also \S\ref{subsec:model_uncert}).

\begin{figure}
\centering
\includegraphics[width=0.49\textwidth]{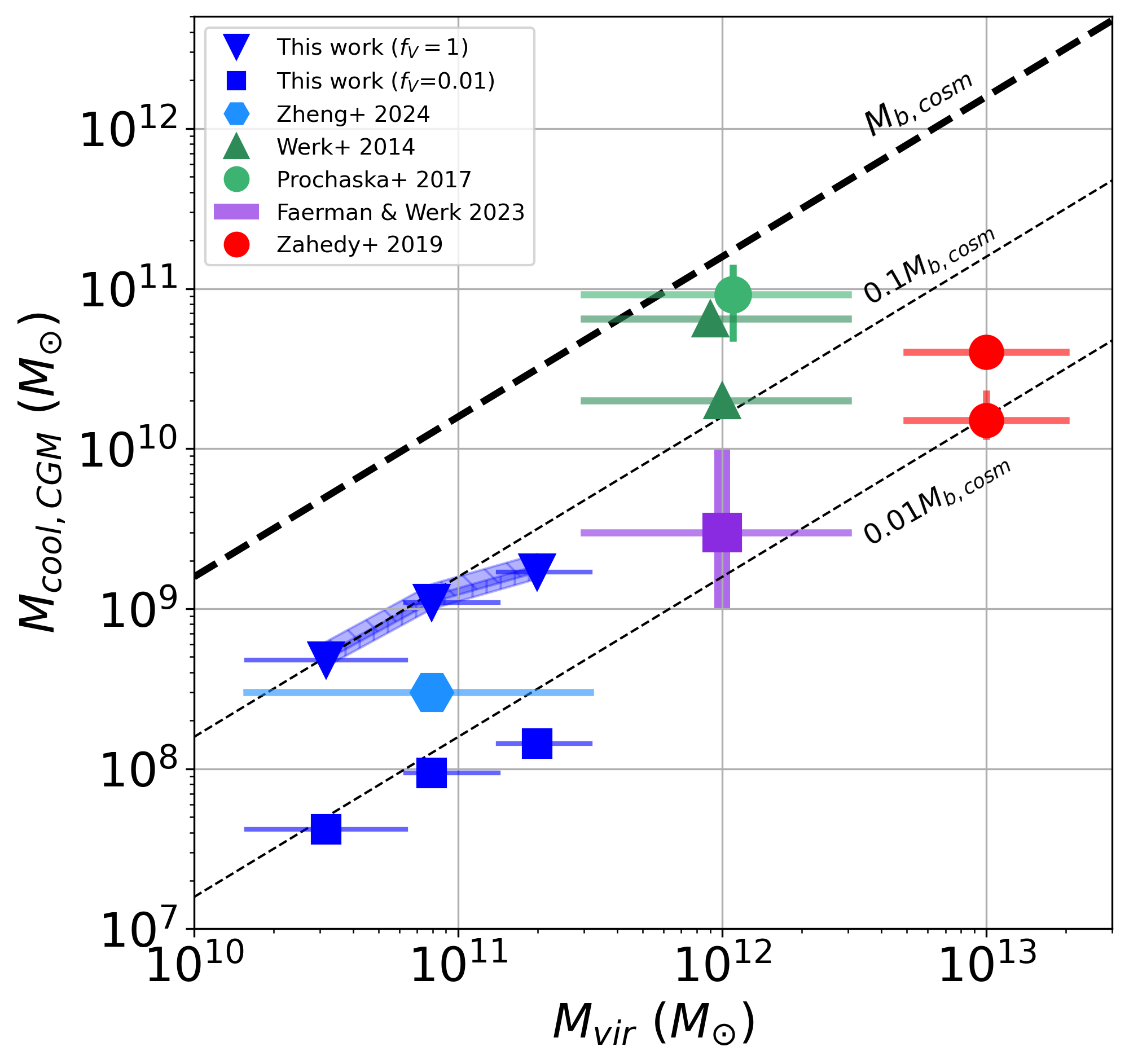}
\caption{Cool CGM mass vs. halo mass. The shaded bands around our upper limits shows the uncertainties due to unknown gas metallicity and redshift range (vertical and diagonal hatching, respectively). Horizontal error bars for our results indicate the halo mass bin widths, and for all other results - the range of halo masses in each sample. Dashed lines show fixed fractions of the halo cosmological baryon mass budget, $\mbcosm = \frac{\Omega_b}{\Omega_m} \m200$. (See \S\ref{subsec:model_results} and \S\ref{subsec:disc_massive}.)
}\label{fig:mass}
\end{figure}

Figure~\ref{fig:mass} shows our results for the cool gas mass as a function of the halo mass, by the blue triangles. The horizontal error bars show the bin widths, and the shaded bands show the uncertainties in our results, discussed in \S\ref{subsec:model_uncert}. The black dashed contours show the masses that constitute fixed fractions of the halo cosmological baryon mass budget, given by $\mbcosm \equiv \m200\Omega_b/\Omega_m  = 0.158 \m200$ \citep{Planck16}. The thick line shows $\mcgmcool = \mbcosm$, and the thin lines are for $10\%$ and $1\%$ of $\mbcosm$.

Our main result is that the observed HI column densities, together with the assumption of volume filling cool CGM, suggest $\mcgmcool \approx 5 \times 10^8-2 \times 10^9$~\msun. Furthermore, these masses constitute $5-10\%$ of the cosmological baryon budgets of the DM halos.

As shown in \S\ref{subsec:model_analytic}, these masses and baryon fractions, inferred assuming volume-filling gas, constitute upper limits. Clumpy cool gas, with $\fv<1$, can reproduce the same HI column densities with higher gas densities and lower masses (see Figure~\ref{fig:thermal_gen}). To demonstrate this point, we fit to the data models with $\fv=0.01$, motivated by studies of MW-mass galaxies (see \S\ref{subsec:disc_massive}). From now on we refer to these as the clumpy models, and they are shown in Figure~\ref{fig:mass} by the blue squares. The gas masses in this scenario are a factor of $\approx 11$ lower than for the nominal models, in good agreement with Equation~\eqref{eq:mass_prop}. The density, temperature, and pressure profiles are shown by the dotted curves in the bottom panels of Figure~\ref{fig:col_data}. Compared to the nominal models, the higher gas densities in the clumpy models result in higher neutral hydrogen fractions, leading to total column densities lower by a factor of $\approx 11$.

\citetalias{Zheng24} presented a slightly different model for the cool gas distribution, with both the density and volume filling fraction varying with radius. The entire data set was modeled using a single halo mass of $\m200 =  10^{10.9}$~\msun~(the sample median), and \citetalias{Zheng24} showed two specific parameter combinations that reproduced the HI column density profile - model A and B (see their Figure~5). Model A had the higher volume filling fractions, $\approx 0.1-1$, and total cool gas mass, with $\mcgmcool \approx 3 \times 10^{8}$~\msun. This gas mass estimate is shown in Figure~\ref{fig:mass} by the light blue hexagon, spanning the entire range of halo masses of our data set. We can compare this result with the masses we infer in the middle mass bin. With an $\left<\fv\right> \approx 0.3$, the mass estimated by \citetalias{Zheng24} is consistent with our upper limits, and similar to the mass of the $\fv=0.1$ model in \S\ref{subsec:model_full}. This demonstrates that models with a spatial variation in the volume filling fraction are consistent with our limit. Figure~\ref{fig:mass} also shows results from the literature for $\mcgmcool$ in more massive galaxies, and we discuss these in \S\ref{subsec:disc_massive}.

\subsection{Model and Result uncertainties} \label{subsec:model_uncert}

We now address three main systematic uncertainties in our inferred cool CGM mass.

First, as noted in \S\ref{subsec:model_full}, the neutral hydrogen fraction depends on the gas equilibrium temperature, \Teq, which weakly depends on the assumed gas metallicity. Higher metallicities correspond to lower gas temperatures, higher HI fractions, and lower gas masses needed to reproduce the same measured $\nhi$. Since in this work we only address the measured HI columns, we do not constrain the gas metallicity, and it is an assumed parameter in our model. To quantify the effect of this assumption on our results, we construct models with metallicities of $Z'=0.1$ and $1.0$, fit them to the data, and compare the cool gas masses to our nominal models, calculated assuming $Z'=0.3$. The masses are within $\pm 10\%$ of the nominal values, and are shown as the shaded vertically hatched band in Figure~\ref{fig:mass}. 

Second, the HI ion fraction is also a function of the MGRF intensity, $\Phi$, and in our analysis we assumed the value at $z=0$. At low $z$, $\Phi$ varies strongly with redshift, approximately as $\Phi \propto (z+1)^{4.5}$ at $z<0.5$ (see \citetalias{Faerman24})\footnote{We also note there is uncertainty in the radiation field spectral shape and intensity at a given redshift, with different models presented in the literature \citep{HM96,HM12,FG09,KS19}. We use the \citet{KS19} field as it has been shown to be consistent with a large number of observational constrains.}. Higher intensity leads to higher ionization and a higher gas mass needed to reproduce the observed HI columns. To estimate the effect of the MGRF on the gas mass due to the galaxy redshift, we repeat our analysis with the MGRF intensity at $z=0.1$ ($z=0.2$), which $50\%$ ($90\%$) of our galaxies lie below, and find that the gas masses needed are $\approx 30\%$ ($\approx 60\%$) higher than at $z=0$. We show the mass increase corresponding to $z=0.1$ in Figure~\ref{fig:mass} by the shaded diagonally hatched band. For masses estimated at $z=0.05$, the mass uncertainty corresponding to $0 \leq z \leq 0.1$ is $\pm 15\%$, similar to the metallicity uncertainty. In the future, larger data sets will allow binning in redshift and minimizing this uncertainty.

Finally, some of the lines of sight in our sample have lower limits on the HI column densities. Our nominal models fit the data assuming the actual values are not far above these limits. To test the sensitivity of our results to this assumption, we replace the lower limits by values that are factors of 3 or 5 higher, and repeat our fits. We find that the factor $3$ increase in HI columns increases the gas mass by $1.6$, $1.4$, and $1.3$ for the low-, mid-, and high-mass bins, respectively, and by $2.0$, $1.7$, and $1.5$ for the factor of $5$ increase. On average, for the full sample, the gas mass increases by a factor of $1.4$ and $1.65$. While these differences are not negligible, even a factor of $2$ increase in $\mcgmcool$ results in $\lesssim 20\%$ of the halo baryon budget in the cool CGM. More importantly, we note that high HI column densities may result from halo substructure or high-density regions, atypical of the average gas density distributions (see also discussion in \citetalias{FW23}). In this case, their contribution to the total gas mass will be much smaller than estimated from including them in a 1D column density profile. Here too, larger data sets will allow a better characterization of the typical HI column densities and outliers.

\section{Discussion} \label{sec:disc}

We now address the implications of our results for gas and galaxy properties. First, we discuss the gas densities and pressures in our models, consider the baryon budget and the metal masses in the CGM, and estimate the accretion rates onto the galaxy. We then compare our results to studies of more massive galaxies, and examine possible contributions to the measured HI column density from the warm/hot CGM and the IGM.

\subsection{Gas densities and pressures} \label{subsec:disc_pressure}

The cool CGM densities in our nominal models are low, with $\nh \sim 10^{-5}$~\cmv~(bottom-left panel of Figure~\ref{fig:col_data}). With $T = \Teq \approx 10^{4}-4 \times 10^{4}$~K, the gas thermal pressures are also low, in the range $P/k_B \approx 0.1-2$~K~\cmv~(bottom-right panel). The blue shaded region in Figure~\ref{fig:IGMcont} shows the range of densities and temperatures for the cool CGM in our models. The top x axis shows the overdensity compared to the cosmic mean, and for our nominal models $n/\bar{n} \sim 10-100$, close to IGM overdensities (see also \S\ref{subsec:disc_igm_cont}).

These nominal low densities are the result of the low measured HI columns, with $\nhi \approx 10^{14}-10^{15}$~\cmc. However, as discussed in \S\ref{subsec:model_results}, the nominal models provide upper limits on the total gas mass and lower limits on the gas density. For lower $\fv$, the gas density increases approximately as $\propto f_{\rm V}^{-1/2}$ (see \S\ref{subsec:model_analytic}). The equilibrium temperature only varies weakly with density, $\Teq \propto n_{\rm H}^{-0.15}$ (see Equation~12 in \citetalias{Faerman24}), and the thermal pressure increases with $\nh$. For example, in our clumpy models, with $\fv=0.01$, the gas densities are higher by a factor of $\approx 11$ compared to the nominal, and the gas pressures are higher by a factor of $\approx 5$ (see dotted curves in the bottom panels of Figure~\ref{fig:col_data}).

How do these pressures compare to the pressure in the warm/hot CGM? Recent works addressed absorption by higher ions in dwarf galaxies \citep[e.g.,][]{Bordoloi14, liang14, Johnson17, Mishra24}. However, the interpretation of these observations is challenging and we still do not have strong constraints on the warm/hot gas physical properties for a comparison between the two phases. We can discuss two possible cases. Densities below $\nh \lesssim 3 \times 10^{-6}~\cmv$ in the CGM are less likely in the warm/hot phase, since these would be similar to IGM densities and very low baryon overdensities (see Figure~\ref{fig:IGMcont}). At densities similar or higher than those of the cool phase (for $\fv \sim 1$) and higher temperatures, the warm/hot CGM will have pressures higher than the cool gas. This is similar to the pressure discrepancy between the phases found by \cite{Werk14} and \citetalias{FW23} in the CGM of MW-mass galaxies, and may suggest some pressure imbalance between the two phases or require nonthermal support in the cool phase. However, for $\fv \ll 1$, the cool phase can coexist in thermal pressure equilibrium with the warm/hot gas (see also \citealp{Zahedy19,Qu23}).

\subsection{Baryon and metal budgets} \label{subsec:disc_budgets}

Our analysis suggests that the cool CGM constitutes $\lesssim 10\%$ of the total halo baryon budget in dwarf galaxies (Table~\ref{tab:params} and Figure~\ref{fig:mass}). The stellar masses in these halos are also low, with mean values of $\mstar \approx 2.5 \times 10^{7}$ (low-mass bin), $2.0 \times 10^{8}$ (mid), and $1.1 \times 10^{9}$~\msun~(high), and $\approx 0.3$~dex dispersion in each bin. These correspond to $M_*/\m200 \approx 0.001-0.006$, or $\approx 1-4\%$ of the halo baryon budget. Together, the stars and the cool CGM account for less than $15\%$ of the halo baryon mass budget in dwarf galaxies. Where are the remaining $\approx 85\%$?

In one scenario, a large fraction of the baryon budget is in the warm/hot, volume-filling CGM, similar to more massive systems (galaxy clusters, and possibly the Milky Way). This phase can be traced by higher ions such as OVI (see \citealp{Johnson17}; \citetalias{Mishra24}). Another option is that in these low-mass systems a significant fraction of the baryons escaped the shallow(er) potential well completely and was ejected into the IGM (see \citealp{Ayromlou23}), returning some of the gas to the cosmic web and enriching it with metals. We now estimate the metal masses and budgets.

In this work, we focus on the HI and do not infer the metallicity of the cool CGM. Furthermore, the absorption measurements reported by \citetalias{Zheng24} consist mostly of upper limits for the metal ions, and the metallicity of the cool CGM is not well constrained. In our calculations, we adopt a nominal value of $Z'=0.3$ and consider a range of $0.1-1.0$. Adopting the solar bulk composition from \cite{Asplund09}, at a solar metallicity metals constitute 0.0142 of the total gas mass, or 0.0144 of the hydrogen and helium mass. The metal mass associated with a given $\mcgmcool$ and metallicity is then given by
\beq\label{eq:mass_metals}
M_{\rm met,cCGM} = 4.3 \times 10^{6} \left(\frac{\mcgmcool}{10^9~\msun} \right) \left( \frac{Z'}{0.3} \right) \msun ~~~.
\eeq
Table~\ref{tab:params} provides the metal masses for the models discussed in \S\ref{subsec:model_results}. These can be used to calculate the column densities of individual metal ions in the cool phase for comparison with future observations (see also \citetalias{Zheng24}).

We can compare these masses to the total metal mass produced in a galaxy over its lifetime. Given the galactic stellar masses and assuming a yield of $0.033$~\msun~of metals per solar mass of stars formed over Hubble time \citep{Peeples14}\footnote{\citet{Telford19} present a fiducial model with a yield of 0.0231, $30\%$ lower than \citet{Peeples14}, and the uncertainties estimated for each of these is higher than the difference between the two studies, with a range of $0.0238-0.0453$ in \citet{Peeples14} and $0.0111-0.0557$ in \citet{Telford19}.}, we estimate that the mass-binned galaxies in our sample produced $M_{\rm met,formed} \approx 8.3 \times 10^5$, $6.6 \times 10^6$, and $3.6 \times 10^7$~\msun. Comparing these to the metal masses in our models gives the fraction of metals in the cool CGM, $f_{\rm met} \equiv M_{\rm met,cCGM}/M_{\rm met,formed}$. For the nominal models, $f_{\rm met}$ varies between $\approx 2$ and $0.2$ for the low- and high-mass bins, respectively. For the clumpy models, the gas and metal masses are reduced by a factor of $\approx 10$, and $f_{\rm met} \lesssim 20\%$. The metallicity of the CGM provides another degree of freedom, with possible values at least $\pm 0.5$~dex from the nominal $Z'=0.3$. We conclude that our calculations do not provide strong constraints on the metal fraction in the cool CGM of dwarf galaxies, and more detailed estimates and predictions are left for future work.

\subsection{Accretion vs. star-formation rates} \label{subsec:disc_accretion}

In classical models of galaxy formation and evolution, the cool CGM falls to the center of the halo within the dynamical time and fuels star formation in the galaxy \citep[e.g.,][]{MB04,Afruni19}. We now estimate the accretion rates suggested by our inferred masses.

The halo dynamical time is given by $\tdyn(r) = \sqrt{2r/g(r)}$. At a given redshift and radius, it is a function of the halo overdensity, independent of halo mass. At $z=0$, \tdyn~increases from $\approx 1$~Gyr at $0.3 \r200$ to $\approx 4$~Gyr at $\r200$. Using the cool gas mass in our nominal models, we can write
\beq
\dmacc \approx \frac{\mcgmcool}{\tdyn} \approx 1.0 \left( \frac{\mcgmcool}{10^9~\msun} \right) \left(\frac{\tdyn}{{\rm Gyr}} \right)^{-1} ~ \msuny ~.
\eeq

The average SFRs in our sample are $0.002$, $0.09$, and $0.16$~\msuny, with a scatter of $\sigma \approx 0.5$~dex in each bin. These rates are lower than our estimated accretion rates for the nominal masses. However, as noted earlier, the latter provide upper limits on \mcgmcool, and the rates for our clumpy models are a factor of $\approx 10$ lower. Hence, we estimate that the accretion rates of the cool CGM are above or similar to the Star formation rate (SFR) in these galaxies, which may allow them to continue star formation at the current rates for the next $\sim$~Gyr or so.

One caveat to this estimate is that it assumes that the cool CGM is bound to the galaxies, and is not supported against infall. While the thermal pressure in the cool phase is not enough to support the gas (since PIE $\Teq$ is lower than the virial temperatures of these galaxies), nonthermal mechanisms such as turbulent motions, magnetic fields, and cosmic rays may be important (e.g., \citealp{Lochhaas23, Ramesh24b,Sultan24} in more massive galaxies). Furthermore, star formation is an inefficient process with only a fraction of the gas mass actually going into stars. For example, if galactic outflows are mass-loaded, accretion rates have to be significantly higher than the SFR \citep[e.g.,][]{Dave11a,Muratov15}. If a nonnegligible fraction of the cool CGM is outflowing or supported by nonthermal pressure, the accretion rates will be reduced, possibly slowing down or completely stopping star formation in these galaxies.

\subsection{Comparison to massive halos} \label{subsec:disc_massive}

We now compare our results for $\mcgmcool$ in low-mass halos to the results reported in the literature for more massive galaxies, and these are shown in Figure~\ref{fig:mass}.

\citet{Werk14} and \citet{Prochaska17} analyzed data from the COS-Halos survey, probing the CGM of $\sim L^*$ galaxies at $z\approx 0.2$, with a typical stellar mass of $\approx 3 \times 10^{10}~\msun$~and halo mass of $\approx 10^{12}~\msun$. Both studies used photoionization modeling of the measured metal columns to translate the HI to total hydrogen column and then integrated over the column density profile, observed out to $\imppar \approx 160$~kpc ($\approx 0.6 \rvir$), to obtain the total (cool) gas mass. \citet{Werk14} inferred two lower limits due to line saturation - a strict limit with $\mcgmcool > 2.1 \times 10^{10}~\msun$~within $\imppar=160$~kpc, and $>6.5 \times 10^{10}~\msun$~within 300 kpc, the mean virial radius for the sample (both shown in Figure~\ref{fig:mass} by the dark green triangles). \citet{Prochaska17} used a larger set of hydrogen transitions for better measurements of the HI column densities and estimated $\mcgmcool(\imppar<160~{\rm kpc}) = 9.2^{+4.3}_{-4.3} \times 10^{10}~\msun$~(shown by the light green circle), consistent with the cool CGM constituting most of the halo baryon budget\footnote{However, we note that the total mass estimate in \citetalias{Prochaska17} is dominated by a few sightlines with very high column densities, which may be atypical for the HI column density profile of the average galaxy in the sample.}. In a recent study, \citetalias{FW23} constructed a multiphase model for the (radial) gas distribution, allowing forward modeling the observed column densities. They presented a nominal model with $\mcgmcool = 3 \times 10^{9}$~\msun~and demonstrated that a range of $10^9-10^{10}$~\msun~(shown by the magenta square and bar), constituting $\sim 1-10\%$ of the baryon budget, reproduces the measured columns of the COS-Halos star-forming galaxies.  In a recent study, \cite{SB24-mcc} fit a misty CGM analytic model to MgII measurements from a wide range of data sets and estimated $\mcgmcool \sim 10^{10}$~\msun.

At even higher halo masses, \citet{Zahedy19} studied the CGM of luminous red galaxies (LRGs), with $M^*>10^{11}~\msun$, at $z \approx 0.4$. The median stellar mass in the sample is $1.6 \times 10^{11}$~\msun, and the corresponding halo mass is $\approx 10^{13}$~\msun. They used the Hubble Space Telescope/COS and ground observations to measure hydrogen and metal transitions, and photoionization modeling to infer the cool gas mass. \citet{Zahedy19} estimated that the CGM of a typical LRG contains  $1.5^{+0.7}_{-0.3} \times 10^{10}~\msun$~inside $\imppar=160$~kpc ($\approx 0.4 \rvir$), or $\approx 4 \times 10^{10}$~\msun~inside 500 kpc, the typical virial radius in the sample. These cool gas masses constitute $\approx 1-3\%$ of the halo baryon budget and are shown by the red circles.

We note that there are some differences between these results for $\mcgmcool$, both in the data they are based on and the analysis methods used. First, the galaxies in each data set have different redshift ranges, and evolution in galaxy and CGM properties may affect any trend with halo mass. Second, they probe different geometries  - the observational studies integrate over the column density profile and obtain the mass enclosed in a cylinder, while this work and \citetalias{FW23} forward model the column densities and report the mass inside a spherical radius. The spatial extent of the data also varies between surveys and datasets. While some of these results can be rescaled or taken as lower limits, for this preliminary comparison we present them as they are reported. Finally, the studies of $L^*$ and LRGs span a range of $\approx 0.5-1$~dex in stellar and halo masses (indicated by the horizontal error bars), without binning.

With these caveats in mind, we can speculate about the behavior of $\mcgmcool$ as a function of halo mass, and identify two interesting possibilities. If the $\mcgm$ of dwarf galaxies is close to the upper limit estimated in this work, and close to the \citetalias{FW23} result for $L^*$ galaxies, the baryon fraction in cool gas seems to be fairly constant across three orders of magnitude in $M_{\rm halo}$, with $f_{\rm b,cool} \approx 1-10\%$. However, if the cool gas mass is lower in dwarfs (clumpy models) and closer to the \citetalias{Prochaska17} value in $L^*$ galaxies, $\mcgmcool$ varies significantly with halo mass, both in absolute mass and its fraction of the halo baryon budget. Future work addressing the modeling differences between data sets and obtaining more data will be able to test these scenarios.

\subsection{IGM/CGM contribution} \label{subsec:disc_igm_cont}

\begin{figure}
\centering{\includegraphics[width=0.45\textwidth]{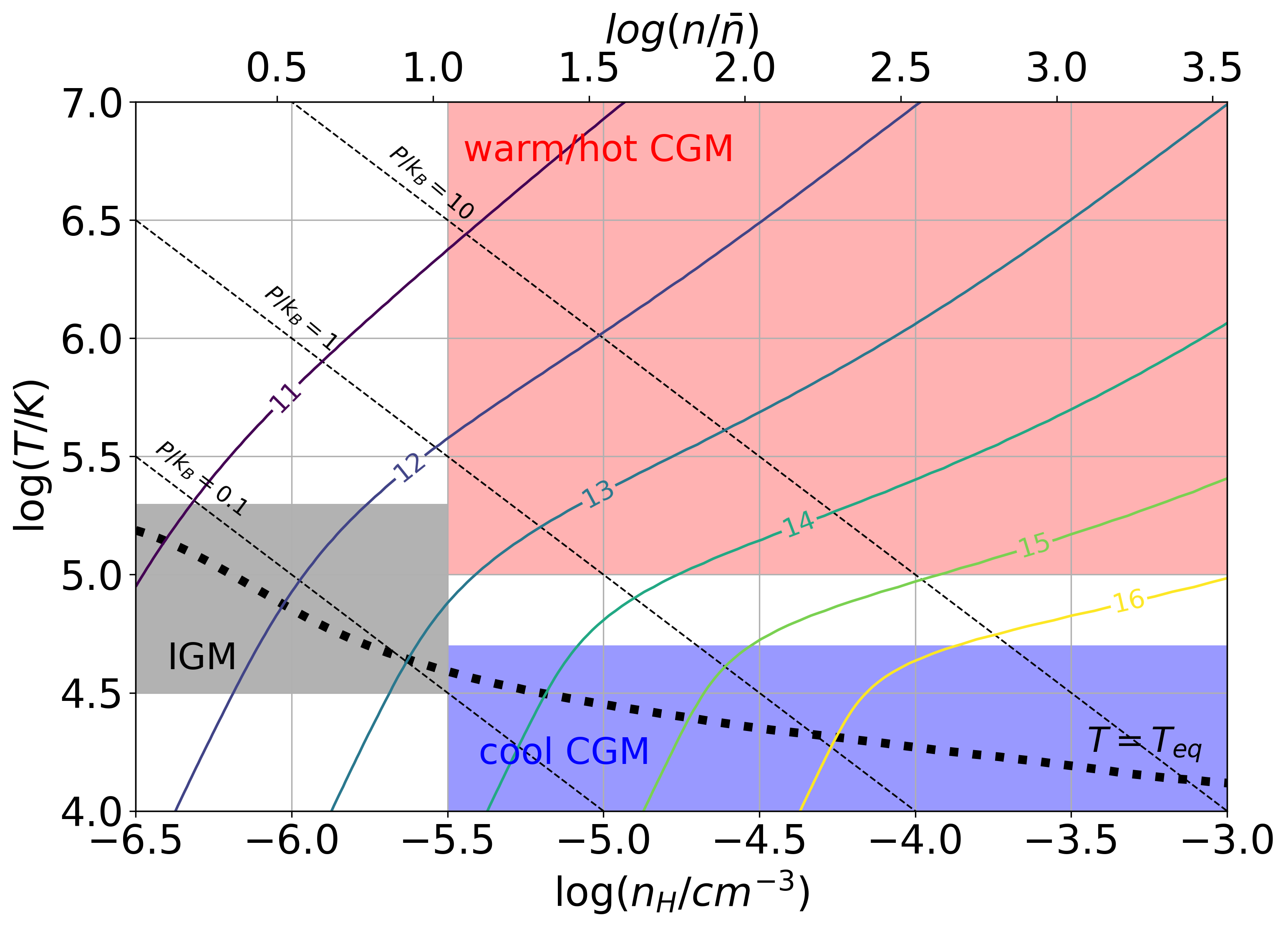}}
\caption{The HI column density for a path length of $150$~kpc~as a function of gas density and temperature (solid contours). Thick dotted curve shows the gas equilibrium temperature, and the thin dashed lines show the gas thermal pressure. The shaded regions show diffuse gas in different regimes. The {\bf cool CGM} (shaded blue) is addressed in \S\ref{sec:model}, and can form fairly large column densities, mainly as a function of gas density (and volume filling fraction). At low densities, the {\bf IGM} (shaded grey) has low HI ion fractions, but with a path length of up to $1$~Mpc can form $N_{\rm HI} \lesssim 10^{14}~\cmc$. The {\bf warm/hot CGM} (shaded red) can span a wide range of temperatures and densities, depending on its formation mechanism (winds vs accretion). Column densities comparable to the observed values ($\approx 10^{14}~\cmc$) can form at relatively low temperatures ($T<3 \times 10^5$~K) or higher densities and pressures (see text in \S\ref{subsec:disc_igm_cont}).} \label{fig:IGMcont}
\end{figure}

In this work, we assumed that the measured HI columns, $\sim 10^{14}$~\cmc~(right panel of Figure~\ref{fig:thermal_gen}), originate in the cool, photoionized phase of the CGM. We now ask how much HI can come from other sources of diffuse gas, and address possible contributions from the warm/hot CGM and the IGM in the vicinity of the galaxy. Figure~\ref{fig:IGMcont} presents our calculation of $\nhi$ as a function of gas temperature and density (bottom x-axis). The top x-axis shows the gas overdensity for a mean baryon density of $\bar{n} = 2.84 \times 10^{-7}$~\cmv, and the thick dotted black line shows the gas PI equilibrium temperature, $\Teq$, both calculated at $z=0$.

The range of properties for the cool CGM is shown by the blue shaded region, with $T \approx \Teq$. The gas densities inferred in this work for volume filling cool gas are at the low end of this range and as discussed in \S\ref{subsec:disc_pressure}, these are lower limits. The IGM is expected to have gas densities lower than the CGM, with overdensities $<10$. For our calculation we assume it is photoionized and at the PIE temperature, similar to the cool CGM, and the approximate properties are highlighted by shaded grey. The warm/hot CGM is expected to have higher temperatures, resulting from accretion-shocked gas or hot galactic outflows \citep{Breitschwerdt00,Birnboim03,faucher-giguere23}, and a possible range of conditions is shown by shaded red.

The solid contours in Figure~\ref{fig:IGMcont} show the HI column forming at a constant gas density and temperature over a $150$~kpc path length, typical of the halo sizes for the dwarf galaxies we address in this work (see \S\ref{subsec:model_data}). The densities in the IGM are low, and the HI column density in the gray region is below $10^{13}$~\cmc. However, the IGM may contribute to the absorption on a larger scale, $\approx 1$~Mpc, producing a column of $\nhi \sim 10^{14}$~\cmv, similar to the measured values (Figure~\ref{fig:col_data}).

The CGM is limited to the halo size, and column densities of $10^{13}$~\cmc~can form at temperatures below $3 \times 10^{5}$~K or densities above $3 \times 10^{-4}$~\cmv. The region where $\nhi > 10^{13}$~\cmc~is roughly defined by $T/n>10^{10}~{\rm K~cm^{3}}$. However, as we go up in density, the gas pressure increases, and isobars for $P/k_B=0.1$, $1$, and $10$~K~\cmv~are shown by the dotted thin lines. Our nominal models have cool gas pressures of $P/k_B \approx 0.1-1$~K~\cmv, and similar pressures in the warm/hot gas only produce $\nhi \lesssim 10^{13}$~\cmc. Nonthermal support in the cool gas may lead to higher total CGM pressures and higher HI columns. Similarly, the clumpy models have densities and pressures higher by a factor of $\approx 10$ and $5$, respectively (see Figure~\ref{fig:col_data}), which allow HI columns up to a few times $\approx 10^{14}~\cmc$.

To summarize, we find that the observed column densities may have some nonnegligible contribution from diffuse gas that is not the cool CGM. Specifically, columns of $\sim 10^{14}~\cmc$ may form in (i) extended IGM on a $\sim 1$~Mpc scale, or (ii) in the warm/hot CGM at temperatures $\approx 3 \times 10^5$~K and densities of $\sim 10^{-4}$~\cmv. The latter scenario requires thermal pressures that may be higher than those in the cool phase.

While the contribution from these other phases can be nonnegligible or even significant, our main result in this work is an upper limit on the mass of the cool CGM. Our assumption that the observed \nhi~forms entirely in the cool gas provides a conservative limit on the gas mass. This mass limit can be adjusted to account for a fraction of the measured column using Equation~\eqref{eq:mass_prop}. The origin of the HI observed around dwarf galaxies is beyond the scope of this paper, and we plan to address this question in future work, with a more detailed model for the multiphase CGM, including a comparison to simulations (see also \citealp{Ho21, Bromberg24}).

\section{Summary}\label{sec:summary}

In this work we study the cool, photoionized CGM of dwarf galaxies, by constructing a simple analytic model for the gas distribution and fitting it to HI column density observations from the literature. We report the first empirical limits on the amount of cool CGM in halos with masses in the range $\m200 \approx 10^{10}-3 \times 10^{11}$~\msun, significantly extending the range of previous works.

For our model, we adopt a power-law density profile, a constant volume filling fraction ($\fv$), and assume that the cool CGM phase is in heating/cooling and ionization equilibrium with the MGRF. We apply the model to the HI column densities measured in UV absorption and reported by \citetalias{Zheng24} and \citetalias{Mishra24}. Our best-fitting models reproduce the observed HI column density profiles (see Figure~\ref{fig:col_data}), and allow us to estimate the cool gas mass, $\mcgmcool$ up to the unknown value of $\fv$.

We show analytically that for a given HI column, $\mcgmcool \propto f_{\rm V}^{1/2}$. Models with volume-filling cool gas ($\fv = 1$), provide an upper limit on the mass (see Figure~\ref{fig:thermal_gen}). For our sample we estimate $\mcgmcool \lesssim 5 \times 10^8-2 \times 10^9$\msun, and that $\lesssim 10\%$ of the total budget of baryons in these galaxies reside in the cool CGM (see Figure~\ref{fig:mass}). We also construct clumpy gas models, with $\fv=0.01$, motivated by previous studies of more massive galaxies, and show that our analytic relation is a good approximation to the numerical results, giving masses a factor of $\approx 11$ lower than models with $\fv=1$.

We address two uncertainties in our results. First, the gas metallicity, affecting the gas equilibrium temperature and HI fraction, is not well constrained. We fit models with metallicities of $Z'=0.1$ and $1$ and find these lead to $\approx 10\%$ variation in the gas mass compared to a nominal value of $Z'=0.3$. Second, the galaxies in our sample have redshifts in the range $0<z<0.3$, with $\left<z\right>_{\rm med} = 0.1$. We model them at $z=0$, and for $z=0.1$ the inferred $\mcgmcool$ is $30\%$ higher.

The gas densities and pressures in our nominal models provide lower limits, with $\nh \propto f_{\rm V}^{-1/2}$. For cool gas that is volume-filling or close to it, the gas densities are $\approx 10^{-5}$~\cmv, barely above those of the IGM. The gas pressures are lower than expected for the warm/hot CGM, which may suggest either pressure imbalance between the phases, or nonthermal support in the cool gas. For lower volume-filling fractions, the thermal pressures are higher and the cool phase may be in thermal pressure equilibrium with the warm/hot gas (Figure~\ref{fig:col_data}).

Our results have important implications in the context of galaxy evolution. First, our upper limits show that the stellar and cool CGM mass constitute together less that $15\%$ of the halo cosmological baryon budget. The remaining $85\%$ can be either in warm/hot CGM or could have been lost to the IGM through stellar feedback processes. Second, assuming accretion of cool gas onto the galaxy on a dynamical time scale, the accretion rates may be enough to maintain star formation at the present measured rates, depending on the actual gas morphology, kinematics, and pressure support.

We then compare our results to cool gas masses from previous studies of more massive halos, and show that the data may suggest a constant fraction of the halo baryons in the cool CGM (Figure~\ref{fig:mass}). However, given the differences in galaxy samples and analysis methods, inferring any trends of $\mcgmcool$ vs. $M_{\rm halo}$ is still highly uncertain and requires additional work. Finally, we show that the diffuse IGM outside these halos and the warm/hot CGM associated with the galaxies can both have contributions to the total measured columns densities (see Figure~\ref{fig:IGMcont}). Our limits on $\mcgmcool$ in this work are conservative in assuming that all the measured HI column originates in the cool CGM.

This study allows, for the first time, to empirically examine the behavior of the cool CGM mass as a function of halo mass. It also highlights the many questions still open. Where are the rest of the baryons in low mass halos and how can we detect them? How many metals are in the CGM and how many were ejected from the halo into the IGM? What physical processes control the properties of the cool CGM, and how does it interact with the galaxy? Future work, combining observations, models, and numerical experiments, will enable us to answer these questions and obtain a better understanding of the CGM and its role in galaxy evolution.


\vspace{0.5cm}

We thank Yuval Birnboim, Christopher Cain, Orly Gnat, Chris McKee, Matthew McQuinn, Daniel Piacitelli, Prateek Sharma, Zhijie Qu, Jessica Werk, and Fakhri Zahedy for discussions during this work and comments on the manuscript, and Nishant Mishra for assistance with their data set. We also thank the anonymous referee for their comments and suggestions that helped improve this work and manuscript. YF is supported by the NASA award 19-ATP19-0023 and NSF award AST-2007012. This work was partially performed at the Aspen Center for Physics, supported by National Science Foundation grant PHY-2210452. YF thanks the astronomy department at Tel Aviv University, where this work was finalized, for their hospitality.

\software{
{\tt matplotlib} \citep{Hunter:2007},
{\tt numpy} \citep{harris2020array},
{\tt pandas} \citep{pandas_2024},
{\tt scipy} \citep{2020SciPy-NMeth},
{\tt Cloudy} \citep{Ferland17,Chatzikos23},
{\tt Colossus} \citep{Diemer18},
{\tt Universe Machine} \citep{Behroozi19}}.



\appendix
\section{Observational data}
\label{app_data}
\restartappendixnumbering

In this appendix we provide the galaxy properties and observations we use in this work, for easier reproduction of our work, or comparison to other models, observations, or simulations. The galaxy stellar masses, redshifts, physical impact parameters, and HI columns shown in Table~\ref{tab:data} are taken from the individual works reporting them, and these are given in the references column. The halo masses and radii and their uncertainties are calculated using the UNIVERSE MACHINE and COLOSSUS tool kits (\citealp{Diemer18,Behroozi19}, see \S\ref{subsec:model_data}).


 \begin{table*}
 \centering
 \caption{Galaxy properties and CGM observations}
 \label{tab:data}
 \hspace*{-3.7cm}
\begin{tabular}{| c | c || c | c | c | c | c | c | c | c | c |}
\toprule
\# & Galaxy ID & $z$ & $log(M^*/\msun)$ & $log(\m200/\msun)$ & $\r200$ (kpc) & $log(N_{\rm HI}/\cmc)$ & $\imppar$ (kpc) & $\imppar/\r200$ & Ref. \\
\midrule
1   & J2339\_z0.08\_24 &  $ 0.083 $  & $6.80 \pm 0.2$ & $10.21 \pm 0.14$  &  $ 73 \pm  8$ & $>15.00$           &  $ 24 $ &  $0.33 \pm 0.03$ & M24 \\
2   & J2245\_z0.13\_74 &  $ 0.135 $  & $7.10 \pm 0.2$ & $10.36 \pm 0.14$  &  $ 78 \pm  8$ & $<13.01$           &  $ 74 $ &  $0.95 \pm 0.10$ & M24 \\
3   & J2308\_z0.16\_57 &  $ 0.160 $  & $7.10 \pm 0.2$ & $10.40 \pm 0.13$  &  $ 78 \pm  8$ & $<13.51$           &  $ 57 $ &  $0.73 \pm 0.07$ & M24 \\
4   & J0333\_z0.20\_38 &  $ 0.204 $  & $7.20 \pm 0.2$ & $10.44 \pm 0.13$  &  $ 78 \pm  8$ & $<12.39$           &  $ 38 $ &  $0.49 \pm 0.05$ & M24 \\
5   & J0114\_z0.30\_64 &  $ 0.303 $  & $7.30 \pm 0.2$ & $10.51 \pm 0.13$  &  $ 76 \pm  8$ & $<13.59$           &  $ 64 $ &  $0.84 \pm 0.08$ & M24 \\
6  &J122815.96+014944.1&  $ 0.003 $  & $7.33 \pm 0.1$ & $10.43 \pm 0.09$  &  $ 93 \pm  7$ & $13.86 \pm 0.01 $  &  $ 70 $ &  $0.75 \pm 0.05$ & LC14 \\
7   & J0333\_z0.16\_41 &  $ 0.164 $  & $7.50 \pm 0.2$ & $10.59 \pm 0.13$  &  $ 91 \pm  9$ & $14.49 \pm 0.20 $  &  $ 41 $ &  $0.45 \pm 0.04$ & M24 \\
8   & J0154\_z0.31\_73 &  $ 0.313 $  & $7.70 \pm 0.2$ & $10.70 \pm 0.12$  &  $ 88 \pm  8$ & $14.21 \pm 0.04 $  &  $ 73 $ &  $0.83 \pm 0.08$ & M24 \\
9   & D9               &  $ 0.139 $  & $7.73 \pm 0.1$ & $10.67 \pm 0.08$  &  $ 98 \pm  6$ & $13.04 \pm 0.07 $  &  $ 84 $ &  $0.85 \pm 0.05$ & J17 \\
10  & J2245\_z0.28\_42 &  $ 0.276 $  & $7.80 \pm 0.2$ & $10.75 \pm 0.12$  &  $ 94 \pm  9$ & $13.79 \pm 0.04 $  &  $ 42 $ &  $0.45 \pm 0.04$ & M24 \\
11  & D1               &  $ 0.123 $  & $7.93 \pm 0.1$ & $10.77 \pm 0.08$  &  $108 \pm  7$ & $15.70 \pm 0.40 $  &  $ 16$  &  $0.15 \pm 0.01$ & J17 \\
12  & J0028\_z0.22\_78 &  $ 0.219 $  & $8.00 \pm 0.2$ & $10.83 \pm 0.12$  &  $104 \pm 10$ & $<12.35 $          &  $ 78$  &  $0.75 \pm 0.07$ & M24 \\
13  & J0119\_z0.28\_89 &  $ 0.285 $  & $8.00 \pm 0.2$ & $10.85 \pm 0.12$  &  $100 \pm  9$ & $14.12 \pm 0.01 $  &  $ 89$  &  $0.89 \pm 0.08$ & M24 \\
14  & 316\_200         &  $ 0.010 $  & $8.02 \pm 0.1$ & $10.78 \pm 0.08$  &  $121 \pm  8$ & $>14.23 $          &  $ 41$  &  $0.34 \pm 0.02$ & B14 \\
15  & J0110\_z0.28\_85 &  $ 0.280 $  & $8.10 \pm 0.2$ & $10.90 \pm 0.12$  &  $104 \pm 10$ & $15.00 \pm 0.01 $  &  $ 85$  &  $0.82 \pm 0.07$ & M24 \\
16  & D2               &  $ 0.161 $  & $8.13 \pm 0.1$ & $10.89 \pm 0.07$  &  $115 \pm  6$ & $15.06 \pm 0.02 $  &  $ 21$  &  $0.18 \pm 0.01$ & J17 \\
17  & J0357\_z0.13\_114&  $ 0.130 $  & $8.20 \pm 0.2$ & $10.90 \pm 0.12$  &  $119 \pm 11$ & $13.30 \pm 0.07 $  &  $ 114$ &  $0.96 \pm 0.09$ & M24 \\
18  & J2308\_z0.10\_91 &  $ 0.097 $  & $8.20 \pm 0.2$ & $10.90 \pm 0.12$  &  $122 \pm 12$ & $<13.17 $          &  $ 91$  &  $0.74 \pm 0.07$ & M24 \\
19  & D4               &  $ 0.138 $  & $8.23 \pm 0.1$ & $10.92 \pm 0.07$  &  $119 \pm  7$ & $14.10 \pm 0.01 $  &  $ 56$  &  $0.47 \pm 0.03$ & J17 \\
20  & J0154\_z0.12\_15 &  $ 0.116 $  & $8.30 \pm 0.2$ & $10.95 \pm 0.12$  &  $125 \pm 12$ & $>15.50 $          &  $ 15$  &  $0.12 \pm 0.01$ & M24 \\
21  & 172\_157         &  $ 0.017 $  & $8.32 \pm 0.1$ & $10.93 \pm 0.08$  &  $135 \pm  8$ & $>14.57 $          &  $ 52$  &  $0.39 \pm 0.02$ & B14 \\
22  & 124\_197         &  $ 0.026 $  & $8.32 \pm 0.1$ & $10.93 \pm 0.08$  &  $134 \pm  8$ & $14.10 \pm 0.03 $  &  $ 101$ &  $0.75 \pm 0.04$ & B14 \\
23  & D7               &  $ 0.092 $  & $8.33 \pm 0.1$ & $10.97 \pm 0.07$  &  $129 \pm  7$ & $13.94 \pm 0.01 $  &  $ 72$  &  $0.56 \pm 0.03$ & J17 \\
24  & 87\_608          &  $ 0.011 $  & $8.52 \pm 0.1$ & $11.04 \pm 0.08$  &  $147 \pm  9$ & $13.92 \pm 0.03 $  &  $ 135$ &  $0.92 \pm 0.05$ & B14 \\
25  & D8               &  $ 0.095 $  & $8.53 \pm 0.1$ & $11.07 \pm 0.08$  &  $139 \pm  8$ & $<11.96 $          &  $ 79$  &  $0.57 \pm 0.03$ & J17 \\
26  & D5               &  $ 0.144 $  & $8.63 \pm 0.1$ & $11.12 \pm 0.07$  &  $138 \pm  7$ & $14.32 \pm 0.03 $  &  $ 57$  &  $0.41 \pm 0.02$ & J17 \\
27  & J0154\_z0.22\_132&  $ 0.218 $  & $8.70 \pm 0.2$ & $11.16 \pm 0.11$  &  $135 \pm 12$ & $<12.71 $          &  $ 132$ &  $0.98 \pm 0.08$ & M24 \\
28  & 329\_403         &  $ 0.013 $  & $8.82 \pm 0.1$ & $11.20 \pm 0.07$  &  $166 \pm  9$ & $<12.40 $          &  $ 105$ &  $0.63 \pm 0.04$ & B14 \\
29  & 135\_580         &  $ 0.010 $  & $8.82 \pm 0.1$ & $11.20 \pm 0.07$  &  $166 \pm  10$& $13.63 \pm 0.05 $  &  $ 120$ &  $0.72 \pm 0.04$ & B14 \\
30  & 257\_269         &  $ 0.024 $  & $8.82 \pm 0.1$ & $11.19 \pm 0.07$  &  $164 \pm  9$ & $13.78 \pm 0.03 $  &  $ 125$ &  $0.76 \pm 0.04$ & B14 \\
31  & D3               &  $ 0.296 $  & $8.83 \pm 0.1$ & $11.23 \pm 0.07$  &  $134 \pm  7$ & $13.88 \pm 0.01 $  &  $ 48 $ &  $0.36 \pm 0.02$ & J17 \\
32  & 322\_238         &  $ 0.012 $  & $9.02 \pm 0.1$ & $11.30 \pm 0.07$  &  $180 \pm 10$ & $14.26 \pm 0.03 $  &  $ 54 $ &  $0.30 \pm 0.02$ & B14 \\
33 &J112418.74+420323.1&  $ 0.025 $  & $9.03 \pm 0.1$ & $11.30 \pm 0.07$  &  $178 \pm 10$ & $13.97 \pm 0.01 $  &  $ 123$ &  $0.69 \pm 0.04$ & LC14 \\
34  & 93\_248          &  $ 0.026 $  & $9.12 \pm 0.1$ & $11.35 \pm 0.07$  &  $184 \pm 10$ & $<12.26 $          &  $ 124$ &  $0.67 \pm 0.04$ & B14 \\
35 &J121413.94+140330.4&  $ 0.064 $  & $9.13 \pm 0.1$ & $11.37 \pm 0.07$  &  $180 \pm  9$ & $14.20 \pm 0.01 $  &  $ 70$  &  $0.39 \pm 0.02$ & LC14 \\
36 &J112644.33+590926.0&  $ 0.004 $  & $9.13 \pm 0.1$ & $11.36 \pm 0.07$  &  $189 \pm 10$ & $14.26 \pm 0.01 $  &  $ 32$  &  $0.17 \pm 0.01$ & LC14 \\
37  & 210\_241         &  $ 0.025 $  & $9.22 \pm 0.1$ & $11.40 \pm 0.07$  &  $192 \pm 10$ & $>14.65 $          &  $ 116$ &  $0.60 \pm 0.03$ & B14 \\
38  & D6               &  $ 0.184 $  & $9.23 \pm 0.1$ & $11.42 \pm 0.07$  &  $168 \pm  8$ & $14.00 \pm 0.01 $  &  $ 63$  &  $0.38 \pm 0.02$ & J17 \\
39  & 70\_57           &  $ 0.034 $  & $9.32 \pm 0.1$ & $11.46 \pm 0.07$  &  $198 \pm  11$& $>14.65 $          &  $ 37$  &  $0.19 \pm 0.01$ & B14 \\
40  & 316\_78          &  $ 0.039 $  & $9.42 \pm 0.1$ & $11.51 \pm 0.07$  &  $205 \pm 11$ & $>14.43 $          &  $ 58$  &  $0.28 \pm 0.02$ & B14 \\
\bottomrule
\end{tabular}
{\raggedright References: B14 - \cite{Bordoloi14}, LC14 - \cite{liang14}, J17 - \cite{Johnson17}, M24 - \cite{Mishra24} \par}
\end{table*}

\section{Analytic expression for the cool CGM gas mass}
\label{app_analytic}
\restartappendixnumbering

For a constant density CGM, the neutral hydrogen column density can be written as
\beq\label{eq:nhi_gen}
\nhi = \nh L \fhi \fv ~~~,
\eeq
where $L$ is the path length along the line of sight. At low densities the neutral fraction can be approximated as a power-law function of the density
\beq\label{eq:fhi_gen}
\fhi \approx f_{\rm HI,0} \left( \frac{\nh}{10^{-2}~\cmv} \right)^\beta ~~~.
\eeq
For gas at a constant temperature of $T=10^4$~K at $z=0$ and densities in the range $10^{-6} < \nh/\cmv < 10^{-2}$, $\beta \approx 1$, and $f_{\rm HI,0} \approx 0.1$, so we can write $\fhi \approx 10\nh$. This approximation is accurate to within $5\%$ at $\nh < 10^{-3}~\cmv$ and $20\%$ close to $\nh=10^{-2}~\cmv$ where the neutral fraction starts to level off toward unity at high densities. Inserting this into Equation~\eqref{eq:nhi_gen} for the column density, we get
\beq\label{eq:hi_col}
N_{\rm HI} \approx 10 L n_{\rm H}^2 \fv ~~~.
\eeq

The total mass of the cool CGM is given by $\mcgm = \frac{4\pi}{3} \bar{m} R_{\rm CGM}^3 \nh \fv$. Solving Equation~\eqref{eq:hi_col} for the gas density or the volume filling fraction and inserting into the expression for mass, we get
\beq
\mcgm \approx \frac{\pi}{15} \bar{m}R_{\rm CGM}^3 \left(\frac{N_{\rm HI}}{L}\right) n_{\rm H}^{-1} \approx \frac{2\pi}{3\sqrt{5}} \bar{m}R_{\rm CGM}^{3} \left(\frac{N_{\rm HI}}{L} \right)^{1/2} f_{\rm V}^{1/2} ~~~.
\eeq
Since $\fv \leq 1$, for a measured HI column density, volume-filling gas leads to an upper limit on the gas mass. Scaling this to typical quantities, and taking $L=2R_{\rm CGM}$ (a good approximation out to $\approx 0.5R_{\rm CGM}$) we get
\beq
\mcgm
\approx 1.2 \times 10^{9} \times \left( \frac{R_{\rm CGM}}{136~{\rm kpc}} \right)^{5/2} \left(\frac{N_{\rm HI}}{10^{15}~\cmc} \right)^{1/2} f_{\rm V}^{1/2} ~\msun ~~~.
\eeq

Scenarios in which the density and volume filling fraction vary with radius affect the relation $\mcgm \propto N_{\rm HI}^{1/2} f_V^{1/2}$ through the integral on the gas distribution.  A nonconstant gas temperature affects this relation through a different relation between the gas density and neutral hydrogen fraction. For example, for a general case in which the neutral hydrogen fraction is given by $\fhi \approx f_{\rm HI,0} n_{\rm H}^{\beta}$, we can write the column as $N_{\rm HI} = n_{\rm H}^{1+\beta} \fv f_{\rm HI,0} L$. Solving for $\nh$ and inserting into the expression for mass gives
\beq\label{eq:mass2}
\mcgm = \frac{4\pi}{3} \bar{m} R_{\rm CGM}^{3} \left( \frac{N_{\rm HI}}{L f_{\rm HI,0}} \right)^{\frac{1}{1+\beta}} f_{\rm V}^{\frac{\beta}{1+\beta}} ~~~.
\eeq
For gas at the heating/cooling equilibrium with the UV background radiation, $\beta \approx 1.15$, resulting in $\frac{\beta}{1+\beta} \approx 0.535$. This has only a small effect on the $\mcgmcool(\fv)$ relation, which our full numerical calculation accounts for.

\bibliography{References}{}
\bibliographystyle{aasjournal}

\end{CJK*}

\end{document}